\documentclass[12pt]{iopart}

\usepackage{epsfig}

\begin{document}

\title[Two photon decay of $\pi^0$ and $\eta$ at finite temperature and density]{Two photon decay of $\pi^0$ and $\eta$ at finite temperature and density}

\author{P Costa$^1$, M C Ruivo$^1$ and Yu L Kalinovsky$^{2,3}$}
\address{$^1$\ Departamento de F\'{\i}sica, Universidade de Coimbra,
P-3004-516 Coimbra, Portugal}
\address{$^2$\ Universit\'{e} de Li\`{e}ge, D\'{e}partment de Physique B5, Sart Tilman, B-4000, LIEGE 1, Belgium}
\address{$^3$\ Laboratory of Information Technologies,
Joint Institute for Nuclear Research, Dubna, Russia}

\ead{pcosta@teor.fis.uc.pt, maria@teor.fis.uc.pt and kalinov@qcd.phys.ulg.ac.be}

\begin{abstract}
A comparative study of the anomalous decays $\pi^0,\, \eta \rightarrow\gamma\gamma$, at finite temperature and at finite density, is performed in the framework of the three--flavor Nambu--Jona-Lasinio. The similarities and differences between both scenarios are discussed. In both cases the lifetimes of these mesons decrease significantly at the critical point, although this might not be sufficient to observe enhancement of these decays in heavy-ion collisions.
\end{abstract}
\vspace{-4mm} 
\vspace{4mm} 

%%%%%%%%%%%%%%%%%%%%%%%%%%%%%%%%%%%%%%%%%%%%%%%%%%%%%%%%%%%%%%%%%%%%%%%%%%%%%%%%%%%%%%%%%%%%%%%%%%%%%%%%%%%%%%%%%%%%%%%%%%%%%%%%%%%%%%%%%%%%%%%%%%%%%%%%%%

The behavior of pseudoscalar meson observables at high density or temperature is a topic that has attracted a lot of attention, having in mind that possible   modifications in the mass spectrum, lifetime and widths  of those mesons could provide signatures for the restoration of symmetries  associated to phase transitions that are expected to take place at high temperature or density \cite{kanaya,rhic}.  

The electromagnetic decays of the neutral pseudoscalar  mesons $\pi^0$ and $\eta$ deserve special attention  and  calculations of such observables in the framework of different models  may be found in the literature \cite{cleo,hashimoto,tklev,tdavid,tpisarski,oka,dorokhov,costaD}. In fact,  the decays  $\pi^0 (\eta)\,\rightarrow \gamma\gamma$ are responsible for the great percentage of photons in the background of heavy-ion collisions, which makes these processes particularly interesting. 
An open question is whether enhancement or suppression of $\pi^0 (\eta)\,\rightarrow \gamma\gamma$  will be found in experiments  of heavy-ion collisions. Although the lifetimes of these neutral mesons are much longer than hadronic time scales and its decays  might  not be observed inside the fireball,  it is nevertheless   important to understand the physics underlying these processes, that   is probably the same of  more complex anomalous mesonic decays ($\omega\rightarrow \pi\pi\pi\,, \omega \rightarrow \rho\pi$), relevant for heavy-ion collisions \cite{tpisarski}.

Temperature effects on the process $\pi^0 \rightarrow \gamma\gamma$ have been object of several studies \cite{hashimoto,tklev,tdavid}  but less attention has been given to the investigation of  $\eta \rightarrow \gamma\gamma$. Since the structure of the $\eta$ exhibits a mixing of strange and non strange quarks, the calculation of temperature and density effects  for this  decay,  which is  more involved than for $\pi^0  \rightarrow \gamma\gamma$, might  carry relevant information concerning the restoration of symmetries. In fact, chiral symmetry shows a tendency to be restored in the non strange sector, but the same is not evident for the strange sector, which makes the study of the electromagnetic $\eta$ decay  particularly interesting. 
Although, in general, the behavior of several observables with temperature is qualitatively similar to that with density, there are fundamental differences between   media with $T\neq 0\,, \rho=0$ and $T= 0\,, \rho\neq 0$. In the first case the phase transition is  probably a smooth crossover and there are threshold effects on the mesons, that dissociate in $\bar q q$ pairs at the Mott temperature; in the second case the phase transition is first order and the threshold effects mentioned above do not generally take place. The question whether these differences will affect significantly the decays $\pi^0 (\eta) \rightarrow \gamma\gamma$ should be clarified.
 
%%%%%%%%%%%%%%%%%%%%%%%%%%%%%%%%%%%%%%%%%%%%%%%%%%%%%%%%%%%%%%%%%%%%%%%%%%%%%%%%%%%%%%%%%%%%%%%%%%%%%%%%%%%%%%%%%%%%%%%%%%%%%%%%%%%%%%%%%%%%%%%%%%%%%%%%%%

We perform our calculations in the framework of the  three--flavor Nambu-Jona-Lasinio model [NJL] \cite{NJL}, including the determinantal 't Hooft interaction that breaks the $U_A(1)$ symmetry, that  has the following Lagrangian: %
\begin{eqnarray}
{\mathcal L\,}&=& \bar q\,(\,i\, {\gamma}^{\mu}\,\partial_\mu\,-\,\hat m)\,q + \frac{1}{2}\,g_S\,\,\sum_{a=0}^8\, [\,{(\,\bar q\,\lambda^a\, q\,)}
^2\,\,+\,\,{(\,\bar q \,i\,\gamma_5\,\lambda^a\, q\,)}^2\,] \nonumber\\
&+& g_D\,\{\mbox{det}\,[\bar q\,(1+\gamma_5)\,q] +\mbox{det}
\,[\bar q\,(1-\gamma_5)\,q]\}. \label{1} 
\end{eqnarray}

For the calculation of the   decays $H \longrightarrow \gamma 
\gamma$ we consider the triangle diagrams for the electromagnetic meson decays \cite{costaD}.
The corresponding invariant amplitudes are given by (the details can be found in Refs. \cite{costaD,costaPRCbr}): 
\begin{eqnarray}
{\tilde{\mathcal T}}_{H\rightarrow\gamma\gamma}(P,q_1,q_2) &=&
i  \int \frac{d^4 p}{(2\pi)^4}\textrm{Tr} \left\{ \Gamma_H S(p - q_1) 
\hat{\epsilon}_1 S(p)  \hat{\epsilon}_2 S(p + q_2) \right\}\nonumber \\ 
&+& \textrm{exchange}. \label{trian}
\end{eqnarray}
As we have shown in \cite{costaD}, there is a good overall agreement between our results for these  quantities, calculated in the vacuum, and the experimental values \cite {pdb}. 

%%%%%%%%%%%%%%%%%%%%%%%%%%%%%%%%%%%%%%%%%%%%%%%%%%%%%%%%%%%%%%%%%%%%%%%%%%%%%%%%%%%%%%%%%%%%%%%%%%%%%%%%%%%%%%%%%%%%%%%%%%%%%%%%%%%%%%%%%%%%%%%%%%%%%%%%%%

In Fig. 1 we plot ${\mathcal T}_{H\rightarrow \gamma\gamma}\,, \Gamma_{H\rightarrow \gamma\gamma}\,,g_{H\gamma\gamma}$, as functions of temperature and density for $\pi^0$ and for the $\eta$. 

%%%%%%%%%%%%%%%%%%%%%%%%%%%%%%%%%%%%
\begin{figure}[t]
	\begin{center}
		\includegraphics[width=0.49\textwidth]{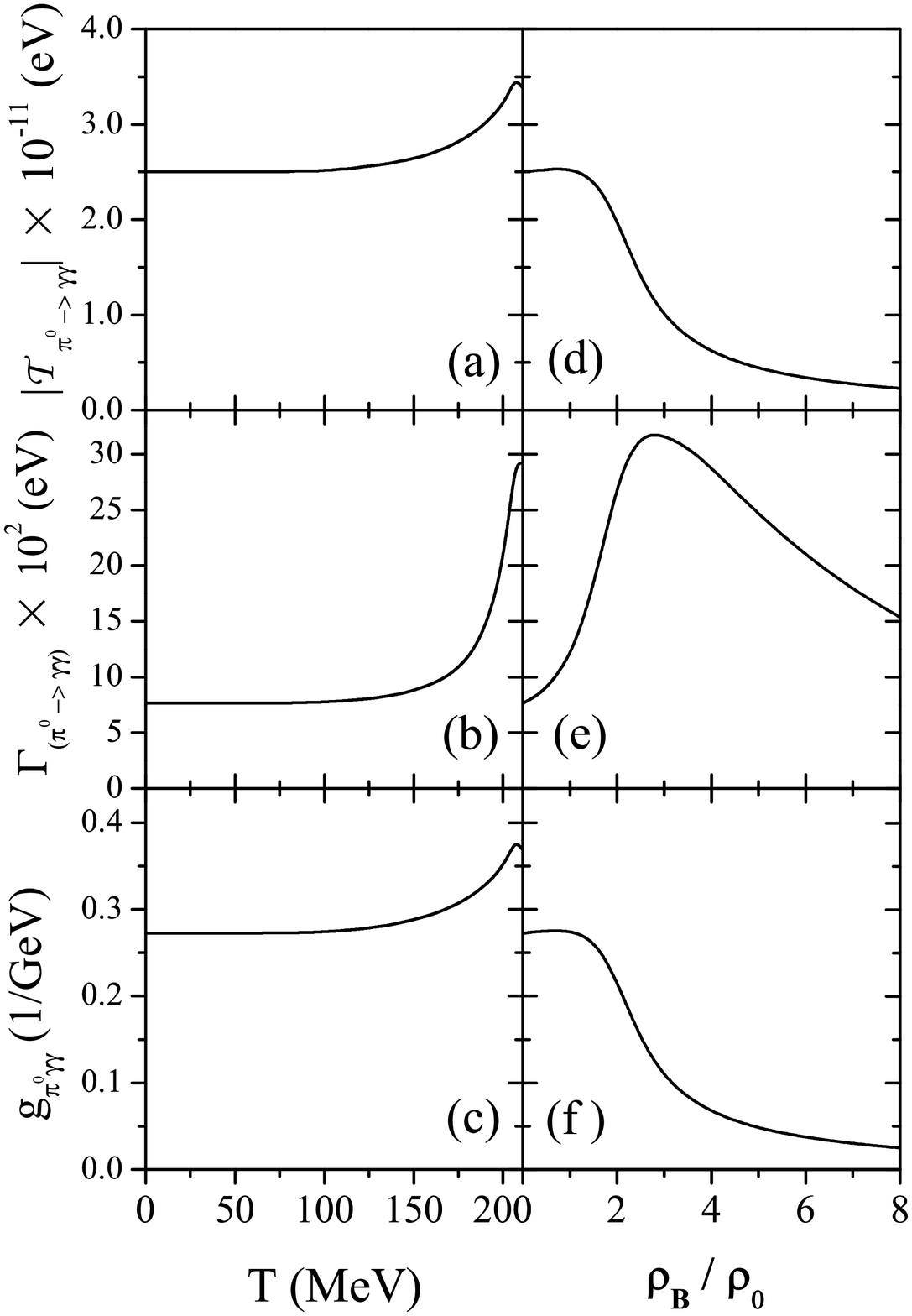}
		\includegraphics[width=0.47\textwidth]{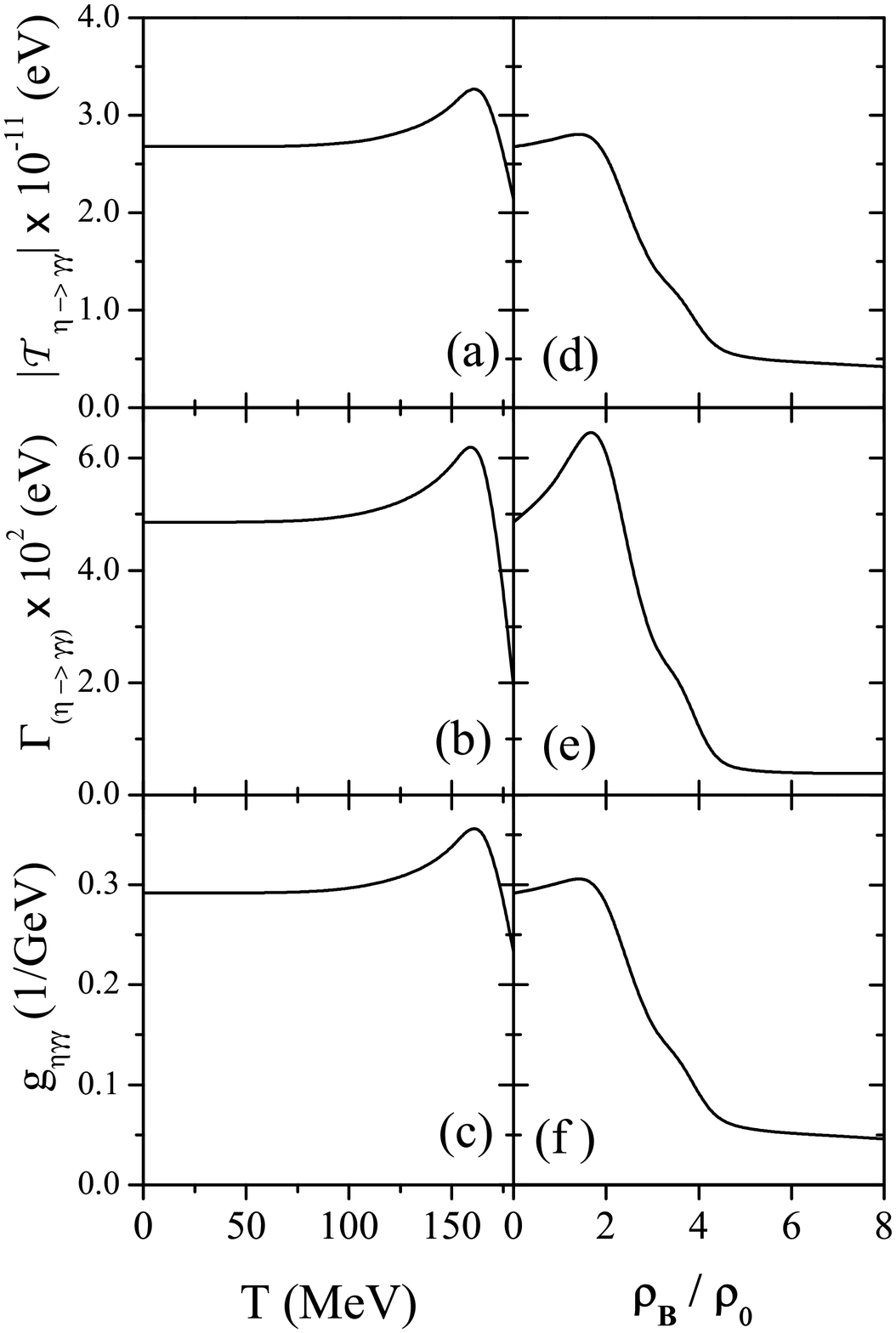}
	\end{center}
	\caption{The decays $\pi^0\rightarrow\gamma\gamma$ (left panel) and 									$\eta\rightarrow\gamma\gamma$ (right panel): 
						transition amplitude, decay width and 
   					coupling constant as functions of temperature (a), b) 
   					and c)), and as functions of the density (d), e) and f)).}
	\label{fig:pi0}
\end{figure}
%%%%%%%%%%%%%%%%%%%%%%%%%%%%%%%%%%%%

We start by analyzing the $\pi^0\rightarrow\gamma\gamma$ at finite temperature and we plot our results only for $ T\leq T^{\pi^0}_{Mott}$,  since above the Mott temperature there is no pion as a bound state, it   dissociates into  $\bar q q$ pairs. Due to this fact, and as discussed in  \cite{hashimoto,tklev,tdavid,costaPRCbr}, at the Mott temperature there are complicated threshold effects leading to an increase of the transition amplitude, as the result of  the balance between the sharp decrease (increase) of quantities  entering in its expression. Concerning the width, and since its expression is of the form
\begin{equation}
\Gamma_{H\rightarrow\gamma\gamma}=\frac{M_{H}^{3}}{64\pi}
|\mathcal{T}_{H\rightarrow\gamma\gamma}|^{2},
\end{equation}
the dominant effect is the increase of the pion mass, leading to 
a sharp increase near the Mott temperature ($T^{\pi^0}_{Mott}\approx212$ MeV).

Concerning the $\eta \rightarrow \gamma\gamma$ decay (Fig. 1, right panel), although qualitatively similar to $\pi^0 \rightarrow \gamma\gamma$, it depends on the strange and non strange  quark masses and related quantities, on the couplings of the meson to the quarks and also on the mixing angle $\theta (M_\eta)$ (see \cite{costaPRCbr}, Fig. 2 and Fig. 1, respectively). 
The behavior of the mixing angle is very smooth, but   threshold effects on the remaining quantities lead to   a behavior similar to the case of the pion. The main difference is that  the peak for the decay amplitude occurs before the Mott temperature for the $\eta$, $T^{\eta}_{Mott}\approx 180$ MeV.

Now let us discuss our results at finite density. We consider the case of asymmetric quark matter with strange quarks in $\beta$--equilibrium \cite{costaI,costaB,costabig}.

Our results for the medium effects on the two photon decay of $\pi^0$,  ${\mathcal T}_{\pi^0\rightarrow\gamma\gamma}\,,\Gamma_{\pi^0\rightarrow\gamma\gamma}$ and $g_{\pi^0\gamma\gamma}$, are plotted in Fig. 1, left panel. 
A meaningful difference between the behavior at $T\not=0\,,\rho=0$ and   $T=0\,,\rho\not=0$ is that, in the last case, although with a weaker coupling to the quarks, the mesons are still bound states, even at high densities.  As a consequence, the sharp threshold effects occurring in the finite temperature case do not take place now, and the behavior of the different quantities is more smooth.
The decrease of ${\mathcal T}_{\pi^0 \rightarrow \gamma\gamma}$, as well as $g_{\pi^0 \rightarrow \gamma\gamma}$ has a simple interpretation, it   reflects the fact that $M_u$ and $g_{\pi^0 \bar q q}$ decrease with density. As it was shown in \cite{costaD}, the decrease of the quark mass is more pronounced than the decrease of $g_{\pi^0 \bar q q}$, in the region $\rho < \rho_c=2.25\rho_0$ the behavior of the transition amplitude is dictated by a compromise between the behavior of the mass and the meson quark coupling, and above $\rho_c$  the mass decrease seems to be the dominant effect. Concerning $\Gamma_{\pi^0 \rightarrow \gamma\gamma}$  it has a maximum at about the critical density, since there are two competitive effects, on one side the decrease of the transition amplitude and, on the other side, the increase of the pion mass. Above the critical density the two photon decay of the pion becomes less  favorable, what reflects that it turns a weaker bound $\bar q q$ pair, as already mentioned before. This is compatible with recent experimental results indicating that pionic degrees of freedom are less relevant at high densities \cite{phenix}.

Concerning the $\eta \rightarrow \gamma\gamma$ decay, although qualitatively similar to $\pi^0 \rightarrow \gamma\gamma$, there are differences that are related to the evolution of the strange quark content of this meson and the behavior of the strange quark in this regime.  
It was show in Ref. \cite{costaB} that  the mixing angle ($\theta=-5.8^0$ in the vacuum) decreases with density, has a minimum  $(\simeq - 25 ^0$) at $\rho\simeq 2.8 \rho_0$,  equals to zero at  $\rho\simeq 3.5\rho_0$, then it increases rapidly  up to the value $\sim 30^0$,  when strange valence quarks appear in the medium ($\rho\simeq 3.8\rho_0$). So, at high densities the  $\eta$ is governed by the behavior of the strange quark mass but, the results for the observables of the $\eta \rightarrow \gamma\gamma$ decay, which are shown in Fig. 1, right panel, are qualitatively similar to those of $\pi^0\rightarrow \gamma\gamma$. The main difference is that the width $\Gamma_\eta \rightarrow\gamma\gamma$ almost vanishes above $\rho=3.8\rho_0$. This behavior of the $\eta$ seems to indicate that the role of the $U_A(1)$ anomaly for the mass of this meson at high densities is less important.

In conclusion, we show that these anomalous decays are significantly affected by the medium, leading to an enhancement of the width around the Mott temperature and $\rho_c$. However,  this enhancement is probably not sufficient to lead to lifetimes shorter than the expected lifetime of the fireball.  Recent experimental results from PHENIX \cite{phenix} show that $\pi^0$ production is suppressed in the central region of Au$+$Au collisions as compared to the peripherical region. This means that $\pi^0 \rightarrow \gamma\gamma$ decay could only be interesting for experimental heavy-ion collisions   at intermediate temperatures and densities. However, although the peak of the $\pi^0$ and $\eta$ widths are at  moderate temperature, the decays are probably observed only after freeze-out, since its life times, $\tau$, are still of the order of $10^{-17}$s and $10^{-18}$s, respectively, much larger than the expected lifetime of the fireball, $10^{-22}$s. A similar conclusion is obtained for matter at finite density and zero temperature, although the behavior of the relevant observables is more smooth.  
 
%%%%%%%%%%%%%%%%%%%%%%%%%%%%%%%%%%%%%%%%%%%%%%%%%%%%%%%%%%%%%%%%%%%%%%%%%%%%%%%%%%%%%%%%%%%%%%%%%%%%%%%%%%%%%%%%%%%%%%%%%%%%%%%%%%%%%%%%%%%%%%%%%%%%%%%%%%

\ack{Work supported by grant SFRH/BD/3296/2000 (P. Costa), by grant RFBR 03-01-00657, 
Centro de F\'{\i}sica Te\'orica and GTAE (Yu. Kalinovsky).
}

%%%%%%%%%%%%%%%%%%%%%%%%%%%%%%%%%%%%%%%%%%%%%%%%%%%%%%%%%%%%%%%%%%%%%%%%%%%%%%%%%%%%%%%%%%%%%%%%%%%%%%%%%%%%%%%%%%%%%%%%%%%%%%%%%%%%%%%%%%%%%%%%%%%%%%%%%%


\section*{References}
\begin{thebibliography}{99}

\bibitem{kanaya}
	Karsch F and Laermann E 
	1994 {\it Phys. Rev.} D {\bf 50} 6954 \\
	Kanaya K 
	1997 {\it Prog. Theor. Phys. Sup.} {\bf 129} 197

\bibitem{rhic} 
	Roland C (PHOBOS Collaboration)  
	2002 {\it Nucl. Phys.} A {\bf 698} 54\\
	Louren\c co C 
	2002 {\it Nucl. Phys.} A {\bf 698} 13

\bibitem{cleo} 
  Gronberg J {\it et al} (CLEO Collaboration) 
  1998 {\it Phys. Rev.} D {\bf 57} 33

\bibitem{hashimoto} 
  Hashimoto T, Hirose K, Kanki T and Miyamura O
  1988 {\it Phys. Rev.} D {\bf 37} 3331

\bibitem{tklev} 
  Klevansky S P  
  1994 {\it Nucl. Phys.} A {\bf 575} 605

\bibitem{tdavid} 
  Blaschke D {\it et al} 
  1995 {\it Nucl. Phys.} A {\bf 592} 561

\bibitem{tpisarski}
	Pisarski R D 
	1996 {\it Phys. Rev. Lett.} {\bf 76} 3084\\
	Pisarski R D, Trueman T L and Tytgat M H G
	1997 {\it Phys. Rev.} D {\bf 56} 7077

\bibitem{oka} 
  Nemoto Y, Oka M and Takizawa M  
  1996 {\it Phys. Rev.} D {\bf 54} 6777\\
  Takizawa M, Nemoto Y and Oka M 
  1997 {\it Phys. Rev.} D {\bf 55} 4083

\bibitem{dorokhov} 
	Anikin I V, Dorokhov A E and Tomio L
	2000 {\it Phys. Lett.} B {\bf 475} 361

\bibitem{costaD}
	Costa P, Ruivo M C and Kalinovsky Yu L  
  2003 {\it Phys. Lett.} B {\bf 577} 129
%  {\it Phys. Lett. }  {\bf B 581} (2004) 274 (E).

\bibitem{NJL}  
	Nambu Y and Jona-Lasinio G 
	1961 {\it Phys. Rev.} {\bf 122} 345\\ 
	Nambu Y and Jona-Lasinio G
	1961 {\it Phys. Rev.} {\bf 124} 246

\bibitem{costaPRCbr}
	Costa P, Ruivo M C and Kalinovsky Yu L   
  2004 {\it Phys. Rev.} C {\bf 70} 048202

\bibitem{pdb} 
  Hagiwara K {\it et al} 
  2002 {\it Phys. Rev.} D {\bf 66} 010001
  
\bibitem{costaI} 
  Costa P  and Ruivo M C 
  2002 {\it Europhys. Lett.} {\bf 60} (3) (2002) 356
%  ,hep-ph/0111301.

\bibitem{costaB} 
	Costa P, Ruivo M C and Kalinovsky Yu L   
  2003 {\it Phys. Lett.} B {\bf 560} 171

\bibitem{costabig} 
  Costa P, Ruivo M C, Kalinovsky Yu L and Sousa C A  
  2004 {\it Phys. Rev.} C {\bf 70} (2004) 025204
%  , hep-ph/0304025.

	
\bibitem{phenix} 
 	Adcox K {\it et al} (PHENIX Collaboratin)
  2002 {\it Phys. Rev. Lett.} {\bf 88} 022301\\
	Adler S S {\it et al} (PHENIX Collaboration)
  2003 {\it Phys. Rev. Lett.} {\bf 91} 072301

%%%%%%%%%%%%%%%%%%%%%%

\end{thebibliography}
\end{document}